\documentclass[11pt]{article}
\usepackage{graphicx} 
\usepackage{
amssymb, 
amsmath,
booktabs,
subcaption,
color, 
soul,
authblk,
setspace,
}

\usepackage[hidelinks]{hyperref}
\usepackage[a4paper, total={6in, 10in}]{geometry}

\begin{document}
\title{Modelling viable supply networks with cooperative adaptive financing}

\author{Yaniv Proselkov\textsuperscript{*}\textsuperscript{†}}
\author{Liming Xu\textsuperscript{†}}
\author{Alexandra Brintrup}
\affil{Department of Engineering, University of Cambridge, The Old Schools, Trinity Ln, Cambridge, CB2 1TN, Cambridgeshire, United Kingdom}
\renewcommand{\thefootnote}{\fnsymbol{footnote}}
\footnotetext{\textsuperscript{*}Corresponding Author: yp289@cantab.ac.uk}
\footnotetext{\textsuperscript{†}These authors contributed equally to this work}

\date{\today}

\maketitle

\begin{abstract}
We propose a financial liquidity policy sharing method for firm-to-firm supply networks, introducing a scalable autonomous control function for viable complex adaptive supply networks. Cooperation and competition in supply chains is reconciled through overlapping collaborative sets, making firms interdependent and enabling distributed risk governance. How cooperative range — visibility — affects viability is studied using dynamic complex adaptive systems modelling. We find that viability needs cooperation; visibility and viability grow together in scale-free supply networks; and distributed control, where firms only have limited partner information, outperforms centralised control. This suggests that policy toward network viability should implement distributed supply chain financial governance, supporting interfirm collaboration, to enable autonomous control.
\end{abstract}


\section{Introduction}
Ongoing international instability means supply risk is a constant challenge for modern, complex supply networks \cite{Goldberg2023IsDeglobalizing}. Rapidly changing global conditions amplify demand and supply fluctuations, making them harder to predict, and causing companies to respond to sudden changes inaccurately. This accumulates up the chain, causing the material bullwhip effect \cite{Chen2000TheEffect}.

Firms use supply chain financing (SCF) to address liquidity gaps caused by mismanagement. However, privileged market positions cause power imbalances between suppliers and buyers,  triggering financial bullwhip effects downstream \cite{Maloni2000PowerChain, Chen2013InternalPerspectives}. Material bullwhips amplify risks, creating bidirectional stressors on mid-tier firms in deep-tier supply networks \cite{Li2020ExploringEffect}. These stressors can limit operational capacity, leading to insolvency and disrupting market conditions for partners and competitors \cite{Ivanov2014TheManagement, Sokolov2016StructuralChain, Dolgui2020DoesChain, Lohmer2020AnalysisStudy, Ivanov2022BlackoutAnalysis}. As partner firms lose customers and suppliers, resultant profit losses cause further bankruptcies \cite{Choi2001SupplyEmergence}. Since modern supply networks are complex adaptive systems (CASs) \cite{Naghshbandi2020ASystems}, these local effects nonlinearly propagate over the system, causing  multidirectional failure cascades, or ripple effects (RE), throughout the supply network \cite{Li2021RippleVulnerability, Dolgui2018RippleLiterature, Dolgui2021RippleDirections}.

Constant variability in supply networks demands sustained monitoring and dynamic responses to emergent conditions \cite{Moura2019Cyber-PhysicalTrends}. While sharing risk-related information among firms is critical \cite{Devadasan2013CollaborativePlanning}, competitive advantages often incentivise keeping data private \cite{Kalaiarasan2022TheFramework}. A privacy-preserving information-sharing framework enables dynamic, decentralised decision-making. This enhances network viability by improving adaptability to shocks, reducing risk, and boosting overall performance \cite{Ivanov2020ViablePandemic}, outperforming centralised control \cite{Abbasi2022SteeringPerspective}.

To define a decentralised decision-making paradigm, it must first be modelled.  Since the network breakdowns that decentralisation is designed to stop are emergent phenomena, arising from complex, multi-agent dynamics, models must have realistic emergence. Real supply networks have this property, but destroying them is impractical. Traditional modelling simplifies systems, removing this property \cite{Hasan2015ModelingChallenges}, so we use complex agent-based network simulation, designing models with adaptive, nonlinear agent interactions, effectively simulating emergent risks.

\section{Literature Review}
\label{sec:litrev}
Supply network infrastructures consist of interdependent assets that produce or relay units to deliver material goods, products, and services \cite{Grafius2020InfrastructureComplexity}. Modern supply networks (SNs) are becoming increasingly complex as diverse needs are met with varied solutions \cite{Harland2003RiskNetworks}, making propagation of risks from external parties \cite{Hofmann2018BigAnalysis, Ferraris2017HowCapabilities} unpredicatble for given firms.
Case studies on infrastructural failures and associated risk mitigation strategies highlight the need for complexity methods to enhance understanding of systemic resilience \cite{Naghshbandi2020ReviewSystems}. Supply network topology significantly impacts risk and supply chain robustness \cite{Nair2011SupplyModel}. Most supply networks follow scale-free complex network distributions \cite{Pathak2007Complexity, Surana2005Supply-chainPerspective}, where degree distributions remain constant across scales, a trait common to many social systems \cite{Barabasi1999EmergenceNetworks}. Random networks are also used, primarily as null models \cite{Erdos1959OnI}.

Modern, multiple supply line supply chains have deep-tier supply chain (DTSC) structure  \cite{Shen2020ProductChain}, exemplified by the Daimler-Chrysler supply chain \cite{Choi2002UnveilingDaimlerChrysler}. This automotive upstream supply chain starts with raw material processors and ends with original equipment manufacturers, forming a resilient ecosystem that uses non-perishable materials recombined into market-ready products. Such supply chains provide valuable test cases for analysing network risk.

A common structure in deep-tier supply chains is the diamond supply topology, characterised by a single first-tier member, many second-tier members, and a single third-tier member. For example, the Honda supply network shows that as one moves from the chain's center to its peripheries, the average number of suppliers per firm decreases \cite{Choi2002UnveilingDaimlerChrysler, Shao2018AIndex}.

 Complex supply structure modelling and analysis benefits supply chain management \cite{Cheng2021FinancingChain, Ledwoch2018TheManagement, Battiston2007CreditNetworks}, particularly to help understand how network structure influences material flow \cite{Dolgui2018RippleLiterature}. Recent reviews highlight advanced artificial intelligence techniques for this purpose \cite{kosasih2024Aapproaches}, demonstrating the value of this through successful implementation in supply chain management \cite{Ashraf2024SupplyManagement}.

When firms exploit power imbalances to delay payments, they create liquidity gaps that financially strain suppliers \cite{Maloni2000PowerChain, Cho2019AnPerformance, Gelsomino2016SupplyReview, Benveniste2001TheLiquidity}. While beneficial in the short-term for the firm imposing delays, this practice can choke suppliers' operational capacity over time \cite{Hofmann2021SupplyHealthy}. These liquidity issues often propagate up the supply chain, with late payments impacting suppliers deeper in the network \cite{Miller2017ThePayments}. Small-to-medium enterprises (SMEs) are particularly vulnerable due to limited resources, weaker bargaining power \cite{Bloom2001RetailerWelfare}, and their positioning between larger firms in deep-tier supply networks \cite{Murfin2014BigThem, Fernandez2019FirmPerformance}.

SCF solutions can address challenges in deep-tier structures, where they function as deep-tier SCF (DTSCF), supporting firms far removed from focal firms \cite{Ogawa2020EvolutionJapan}. While simulation models unifying supply chain and financing dynamics exist \cite{Yang2023AnContagions}, they lack a deep-tier focus. This gap was addressed in \cite{Proselkov2024FinancialModel}, which incorporated power differential-induced risks into deep-tier network simulations. However, the mitigation strategies were limited to agent-specific information, with no visibility into the broader supply chain.

Firm relationship structures shape mutual influence and power dynamics, forming the basis of nexus theory \cite{Yan2015APerspective}. Using centrality metrics on these structures, we use nexus theory to determine nexus risk, the risk a firm poses to the network based on its topological importance \cite{Beaumier2023Cross-NetworkChain, Inomata2024MeasuringFrequency, Rivera-Gonzalez2023Supply-chain-focusedAreas}. Four key centrality measures highlight distinct aspects of ``nexusness'' \cite{Shao2018AIndex}: 
\begin{itemize}
    \item \textbf{Degree centrality} reflects a firm's influence over its partners \cite{borgatti2009oncontext}.  
    \item \textbf{Betweenness centrality} captures firm roles as  brokers of information and materials \cite{Kim2011StructuralApproach}.  
    \item \textbf{Eigencentrality} indicates a firm's provision of indirect links to critical players \cite{Yan2015APerspective}.  
    \item \textbf{Closeness centrality} measures firm access to network information \cite{Kim2011StructuralApproach}.  
\end{itemize}

Topologically important suppliers are nexus suppliers. This means that ripple effect impact is conditioned for by susceptibility to financial squeeze, so nexus effects mean the riskiest firms are the least capable of managing risk \cite{Li2021OutNetworks, Choi2021ManagingChains}. A buyer pressuring a supplier into financial strain risks defaulting the supplier, typically indicating the supplier's lack of critical importance, high competition, and insufficient bargaining power to influence payment terms \cite{Malak-Rawlikowska2019FarmersAssessments}. 

This reflects strong supply chain embedding and high connectivity, leaving weak firms vulnerable to market changes and correlated risks. This often leads to unidentified nexus suppliers \cite{Brintrup2018SupplyField, Ledwoch2018TheManagement, Arora2021HowStudy}, whose critical yet unrecognised positions make them more susceptible to financial stress and bankruptcy. This stress propagates through the network, triggering the bullwhip effect \cite{Proselkov2023FinancialModel}. Therefore, addressing nexusness in risk mitigation strategies can reduce both bullwhip and ripple effects.

The bullwhip effect arises when sudden shifts in product demand lead suppliers to overestimate future demand. To avoid missing a surge, they order excess materials \cite{Lee1997TheChains}. This amplification repeats at each tier, moving upstream and leaving suppliers overstocked. These disruptions impact supply and demand dynamics throughout the supply chain. Companies attempt to predict ecosystem responses, but inaccuracies compound at each tier, intensifying shocks caused by the bullwhip effect \cite{Lee2004InformationEffect}.

To avoid bullwhips, maintain liquidity, and preserve supply chains, firms secure financing from sources like bank loans and SCF provided by specialist fintechs \cite{Gatti2005AFragility}. Bank loans, often the first choice, offer favourable terms and straightforward structures \cite{Cunat2007TradeProviders}. SCF is utilised when bank financing is unavailable, with brokers analysing invoice data to understand network structures, making it a promising source of dynamic nexus information \cite{Leuschner2023ToFinancing}. 

A prior study \cite{Proselkov2023FinancialModel} modelled cash-constrained bank and SCF dynamics using network simulations but did not incorporate nexus information into risk mitigation. Additionally, the models were initialised as lattices, limiting the practical applicability of the insights.

Inappropriate SCF policies can amplify ripple effects. Upstream material bullwhips cause liquidity gaps that demand financing \cite{Chen2016AnEffect}. Financing policies, based on cash flow predictions subject to bullwhip principles, lead to a financial bullwhip propagating downstream \cite{Allen2000FinancialContagion, Cassar2002ContagionNetworks, Ozelkan2009ReversePricing, Chen2013InternalPerspectives}. Thus, when a ripple begins, it propagates through a riskier system and is amplified.

Dynamic financing methods exist \cite{Lee2023DynamicFinTech} but lack network integration. Network integration improves visibility, which is critical for mitigating ripple effects, a major supply chain risk \cite{Sokolov2016StructuralChain}. However, a significant challenge is the conflict between network-integrated financing and the privacy requirements of stakeholders. These mechanisms often require firms to share proprietary information with entities outside direct partnerships, highlighting the need for privacy-preserving solutions.

Privacy preservation provides practical advantages, such as maintaining competition among firms, making the undermining of confidentiality undesirable \cite{Gofman2020ProductionDestruction}. Decentralised financing methods have been proposed to enable transparent supply chain management while respecting privacy \cite{Alirezaie2024DecentralizedFinancing}. The value of independence and distribution is evident in the contrast between ``Fordist'' vertical integration, which fully incorporated supply chains, and modern vertical reintegration exemplified by Apple. Apple enforces its standards while maintaining independent suppliers, achieving superior flexibility and efficiency compared to Ford's model, making it more effective in today's contexts \cite{Moriahra2020VerticalSystem}. Efforts to forecast supply chain dynamics without shared information include privacy-preserving learning using federated approaches \cite{Ali2017SupplyShared, Zheng2023FederatedPrediction}. However, this still involves submitting proprietary information to an entity, which can be problematic for traditionalist managers who may prefer sharing only non-proprietary data.

Local policy-sharing solutions enhance coordination in infrastructural networks, improving functionality across multiple dimensions \cite{Bodo2021Decentralisation:Perspective}. In supply chains, this requires inter-firm information sharing to infer extended partner structures and achieve greater supply chain visibility. However, this is difficult due to challenges such as budget constraints, differing data standards, privacy concerns, conflicts of interest, skill requirements, and the inherent complexity of supply chains \cite{Kalaiarasan2022TheFramework}. As a result, smaller, financially constrained firms — those who would benefit most — are the least equipped to implement such solutions.

Local policy sharing can inform early warning systems to mitigate risks in complex adaptive systems by integrating network analytics with network control \cite{Birge2023DisruptionNetworks, Katehakis2020DynamicItems}. However, this is challenging because many agents in infrastructural networks have limited scopes, and at high levels of network complexity or operational speed, centralised control becomes unscalable and inefficient. A scalable alternative is decentralising control, where agents act as sensors of network dynamics and collectively contribute to system analysis \cite{Schmitt2015CentralizationDisruptions}. This approach remains computationally efficient if each agent limits its sensor range and control scope, preventing any single agent from experiencing excessive computational demands. This has been studied in telecommunications \cite{Herrera2021MiningNetworks}, whereby limiting information to relevant regions of operation and using centrality weighted by data packets held at nodes in these regions to approximate criticality, it was shown that decentralisation well approximates systemic criticality dynamics \cite{Proselkov2020DistributedNetworks, Proselkov2022TheComputationb}. Such investigations have yet to be conducted for supply chain financing.

Limiting visibility facilitates intertwined supply chains \cite{Ivanov2021SupplyStrategies}, fostering cooperation among agents and enhancing resilience to shocks. Nexus-based policy sharing may allow incomplete information sharing while protecting proprietary data. Bayesian network systems have been used to model interdependencies with nodes as variables and edges as interaction probabilities \cite{Hosseini2016ModelingPorts}, but these directly model intertwinement rather than its emergence from organisational structures, as in reality  \cite{Zhu2024EffectsSimulation}. When network structure is unknown but dynamics are understood, Bayesian models are suitable \cite{Troffaes2014AModels}, whereas modelling emergence requires complex network dynamical systems with broader data and simpler interactions \cite{Pagani2019ResilienceNetworks, Feyzmahdavian2012ContractiveLaws, Schafer2018DynamicallyGrids}. Emergent properties can be studied using agent-based modelling, simulating agent interactions to analyse topological evolution and dynamic network behaviour \cite{Yu2016DistributedSystems, Proselkov2022TheComputationb, SalvadorPalau2019Multi-agentPrognostics}. This approach enables modelling complex adaptive supply networks (CASNs) and capturing ripple dynamics \cite{Proselkov2024FinancialModel, Li2021RippleVulnerability}, though long-term survivability remains unmeasured.

This context allows for the analysis of long-term behaviours, addressing systemic risk through viability, which measures dynamic adaptivity and improvement. For growing economies, viability is the key determinant of long-term effectiveness \cite{Ivanov2020ViabilityOutbreak}. This contrasts with performance indicators like resilience, robustness, and efficiency: efficiency reflects current operational performance; robustness is resistance to shocks; and resilience is the capacity to recover to a prior state after a shock \cite{Pettit2010EnsuringFramework, Callaway2000NetworkGraphs}. These measures are event-focused and assume a balanced state from which deviation occurs, whereas viability accounts for ongoing adaptivity.

Supply networks perform optimally near the bifurcation state but exist in finite time \cite{Nawrocki2014AReturns}, meaning stability achievable in the asymptotic limit is unattainable. Consequently, inferring long-term viability dynamics from finite-time simulations is essential for measuring systemic viability. While topological measures have been applied \cite{Liu2022ModelingPandemic}, survival time is proposed as the most suitable metric \cite{Ivanov2023IntelligentViability}, requiring temporal data or simulations, highlighting the utility of agent-based modelling.

\section{Knowledge Gaps and Research Design}
\label{sec:k_gaps}
The challenge lies in understanding the behaviour of scalable, autonomously controlled supply networks to mitigate ripple effect risk. Enhancing visibility into deep-tier network segments offers a potential solution but introduces the unresolved issue of preserving privacy within a scalable framework.

We propose local policy information sharing, anonymised through aggregation via local network topology or partner structure using a graph function. By limiting information sharing to local firms, proprietary data exposure is reduced, and fewer entities access the information.

This highlights a research gap in developing a large-scale, agent-based network simulation environment with a distributed, information-limited, topological risk mitigation mechanism against ripple effects. Such an environment would enable the study of scalable autonomous control, particularly in deep-tier network visibility, and help solve the problem of defining scalable, privacy-preserving nexus methods. It requires simulations starting as complex networks to explore nexus-based risk mitigation while accounting for network nonlinearities.  Data anonymisation could address this gap and offer practical benefits if designed effectively.

We propose to make firm risk mitigation strategies depend on policies of other firms, which would solve the above problems. If all clusters overlap with at least one other cluster,  fully covering the supply network, then all firms implicitly make decisions using information about the whole network, thus this would be an effective ripple effect risk management tool. 

We hypothesise that in heterogeneous topologies, risk can be effectively mitigated through the use of local clusters of limited size, achieving scalability. To test this, we analyse the relationship between the diameter of information clusters — named visibility — and systemic viability.

We run iterations of a complex agent-based network simulation under varying agent visibility scenarios. We assess the outcome space using diverse analytical techniques, proposing novel measures of long-term and indefinite viability based on the simulation results.

Our control function generates collaborative intelligence with limited visibility, specifically designed to mitigate critical failures. Using simulation-based approaches, we study survivability, which is most impacted by this function, directly relating to viability. These simulation models incorporate systemic dynamics, leading us to develop the methodologies: the Long-Term Viability Measure (LTVM); and the Indefinite Viability Measure (IVM).

We evaluate the effectiveness of our proposed collaborative intelligence control method \cite{Devadasan2013CollaborativePlanning} using a dynamic complex adaptive systems model \cite{Choi2001SupplyEmergence} of SCF \cite{Proselkov2023FinancialModel}. Testing over various complex network topologies, parameterisations, and nexus measures, we investigate the relationship between visibility — determined by network structure — and viability, as measured by incidence of long-lived supply chains.

Our goal is to develop a scalable, autonomously controlled supply network for ripple effect mitigation by equipping each firm with information about a limited number of subjectively relevant firms, characterising it through its relationship between visibility and viability.

\section{Methods}
\label{sec:methods}
To conduct our investigation, we modify the simulated environment from \cite{Proselkov2023FinancialModel}, where a network, $G=(V,E)$ of suppliers interconnects acyclically. Connected to the nodes without out-degree is a raw materials node, $r$, and to those without in-degree, a market node $m$. From $m$, every timestep $t$, a quantity of demand is generated by the market node according to a Poisson distribution, such that the demand at timestep $t$ is defined as $d^m_t \sim \text{Poisson}(\lambda)$. A node $u$ upstream of $m$, such that $(m,u) \in E$, is randomly selected according to its market share, $\mu_u$, and that demand is propagated upstream to $u$. An example of this is shown in Fig. \ref{fig:er_graph}. The market share of $u$ is based on its power, $\mathcal{P}_u = \{s', m', l'\}$, or ``small'', ``medium' The node $u$ has a quantity of available stock, $s^u_t$. A purchase order is made by the market equal to $p_m^t = \min(d^m_t, s^u_t)$.

A buyer $u$ uses the relative power between $u$ and seller $v$ to determine how many timesteps $\tau_{uv}$ before $u$ pays $v$ its owed cash amount, stored for $v$ as a receivable $R_v^{t+\tau_{uv}}$, and for $u$ as a payable $P_u^{t+\tau_{uv}}$. The relationship of $\mathcal{P}$ and $\tau$ is shown in Table \ref{tab:2}, following invoice convention \cite{Perko2017Behaviour-basedEvaluation}. Between a firm and $m$ or $r$, there is no delay, so for any $u,v \in V$, $\tau_{mu} = \tau_{vr} = 0$.
\begin{table}
    \centering
        \begin{tabular}{ll|lll}
        && \multicolumn{3}{c}{$\mathcal{P}_{u}$ }\\
        & & $s'$  & $m'$  & $l'$   \\
        & $s'$ & 60 & 90 & 120 \\
        $\mathcal{P}_{v}$ & $m'$ & 30 & 60 & 90  \\
        &$l'$ & 30 & 30 & 60 
        \end{tabular}
    \caption{Table of invoice terms where buyer $u$ with power $\mathcal{P}_{u}$ pays supplier $v$ with power $\mathcal{P}_{v}$ exactly $\tau_{uv}$ days late after $v$ makes a delivery to $u$.}
    \label{tab:2}
\end{table}

Each edge, including those attached to $m$ or $r$, has a material unit value parameter, $s_{uv}$, such that for a sale from $u$, it gets cash $c_u^t= p_v^t s_{uv}$, added to its receivables, updating with $R_v^{'t+\tau_{uv}}  = R_v^{t+\tau_{uv}} + c_u^t$, while losing stock equal to the purchase order. The buying firm loses the same amount of cash, added to its payables with $P_v^{'t+\tau_{uv}} = P_v^{t+\tau_{uv}} + c_u^t$. The market node, $m$, has infinite cash, while the raw materials node, $r$, has infinite stock. 

Nodes use their cash to make upstream purchases, cover fixed operative expenses $o$, and pay fixed delivery costs $e$ per timestep per unit sold, thus making costs scale with activity levels. At the next timestep, any remaining demand at $u$ is propagated upstream to a supplier $v$ following the same rules. Nodes source materials by placing orders with the raw materials node $r$.

A firm, $u$, is removed from the system when its current liquid cash, $\kappa_u^t = R_u^0 - P_u^0 \leq 0$, and total receivables are less than its total payables, where $\sum_{t=0}^{120} R_u^t \leq \sum_{t=0}^{120}P_u^t$, calculated until 120 days in the future. When a firm is removed, all its edges are deleted, and relative market shares of competing firms are recalculated from remaining firms. Every timestep, we update receivables and payables such that $R_u^{'t} = R_u^{t+1}$ and $P_u^{'t} = P_u^{t+1}$, modelling the passing of a single day.

To pay partners and cover operational expenses while avoiding bankruptcy, firms may seek external financing, increasing liquidity in exchange for long-term cash flow. One option is bank financing, constrained by debt $\rho_u^t$ and prior cash position times banking mandate, which depends on firm power. When a firm receives bank financing $b_u^t$, it is added to $\kappa_u^t$ and $\rho_u^t$ (where $\rho_u^0 = 0$), with $b_u^t$ limited by
\[
\rho_u^t + b_u^t  \leq \begin{cases}
    2 \kappa_u^{t-1}, &s_u = s;\\
    3 \kappa_u^{t-1}, &s_u = m;\\
    4 \kappa_u^{t-1}, &s_u = l.
\end{cases}
\]
Bank financing cuts future earnings to increases current liquid cash. This cost is expressed as firms repaying loans, with interest $\beta$, added as the payable, $P_u^{\prime t+120} = P_u^{t+120} + (1 + \beta)b_u^t$. To minimise long-term cash reduction, firms seek the smallest financing amount required to bring current cash $\kappa_u^t$ to a threshold $f_u^t$ — a safety cash stock. Our distributed collaborative intelligence control method is applied to determine the financing threshold.

If bank financing isn't enough due to the debt limit, where $\kappa_u^t + b_u^t < f_u^t$, then a node uses SCF. It reduces receivables a fixed time in the future $\iota$ to increase liquid cash by that amount, discounted by a factor of $\gamma$, where $\kappa_u^{'t} = \kappa_u^{t} + s_u^t$, where $s_u^t \leq R_u^{t+\iota} (1 - \gamma)$ and $R_u^{'t +\iota} = R_u^{t+\iota} - s_u^t(1 + \gamma)$.

\subsection{Complex Network Topology} \label{sec:comnettop}
\begin{figure}
    \centering
    \includegraphics[width=0.75\linewidth]{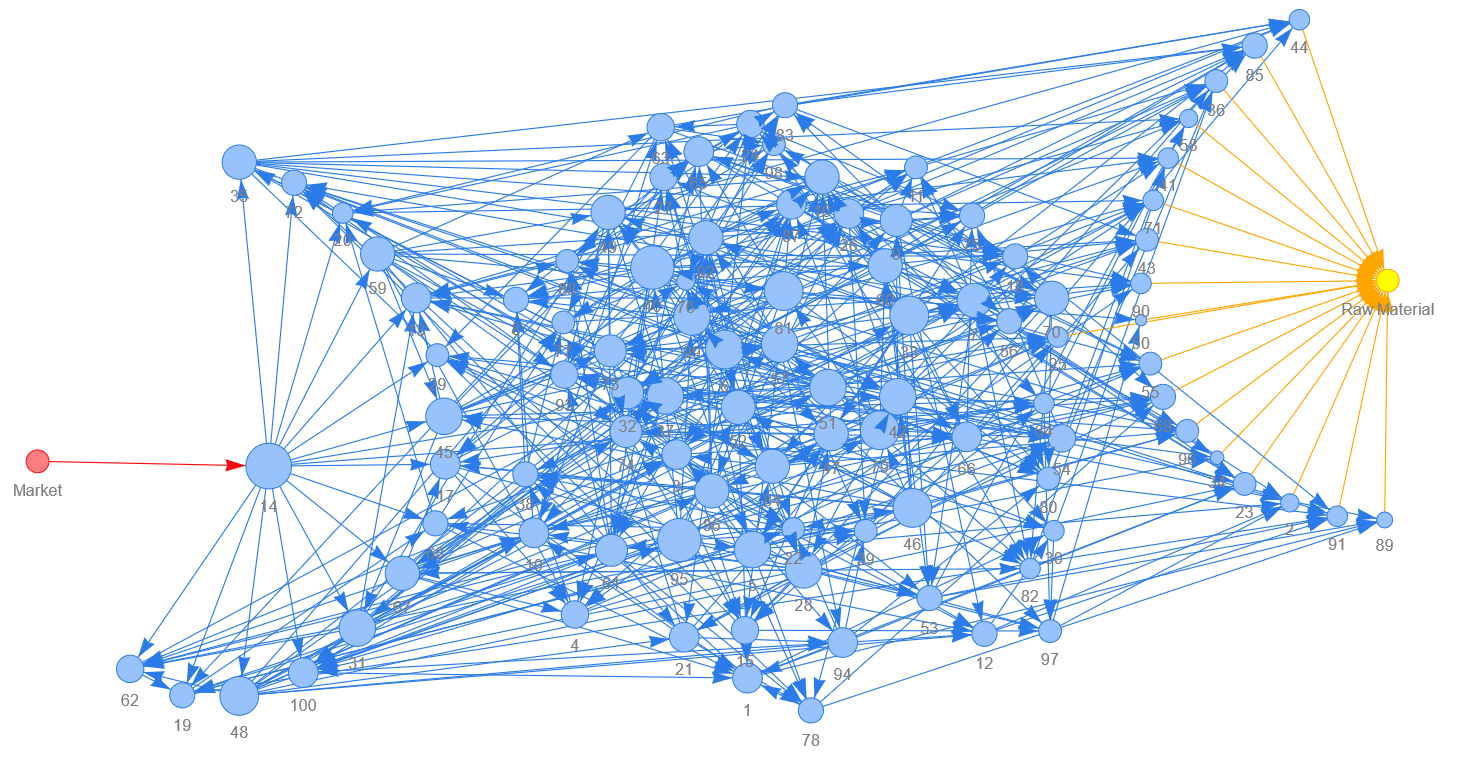}
    \caption{$E(100,0.1), \delta = 7$ supply network: A visualisation of a random, Erdo\H{o}s R\'{e}nyi, $E(n,p)$ graph topology with structural modifications to represent a deep tier supply chain. $n=100$, $p = 0.1$, and diameter, $\delta = 7$.}
    \label{fig:er_graph}
\end{figure}

To construct network topology for study in a complex deep tier network we use the $G(N,p)$ Erdős–Rényi (ER) random graph  model, which generates $G$ according to the fixed number of nodes $N$ and a fixed likelihood $p$ of an edge $uv$ existing between any two nodes \cite{Erdos1959OnI}, visualised in \ref{fig:er_graph}. We treat this as a base case, capturing a long chain with complex unstructured design. 

\begin{figure}
    \centering
    \includegraphics[width=0.7\linewidth]{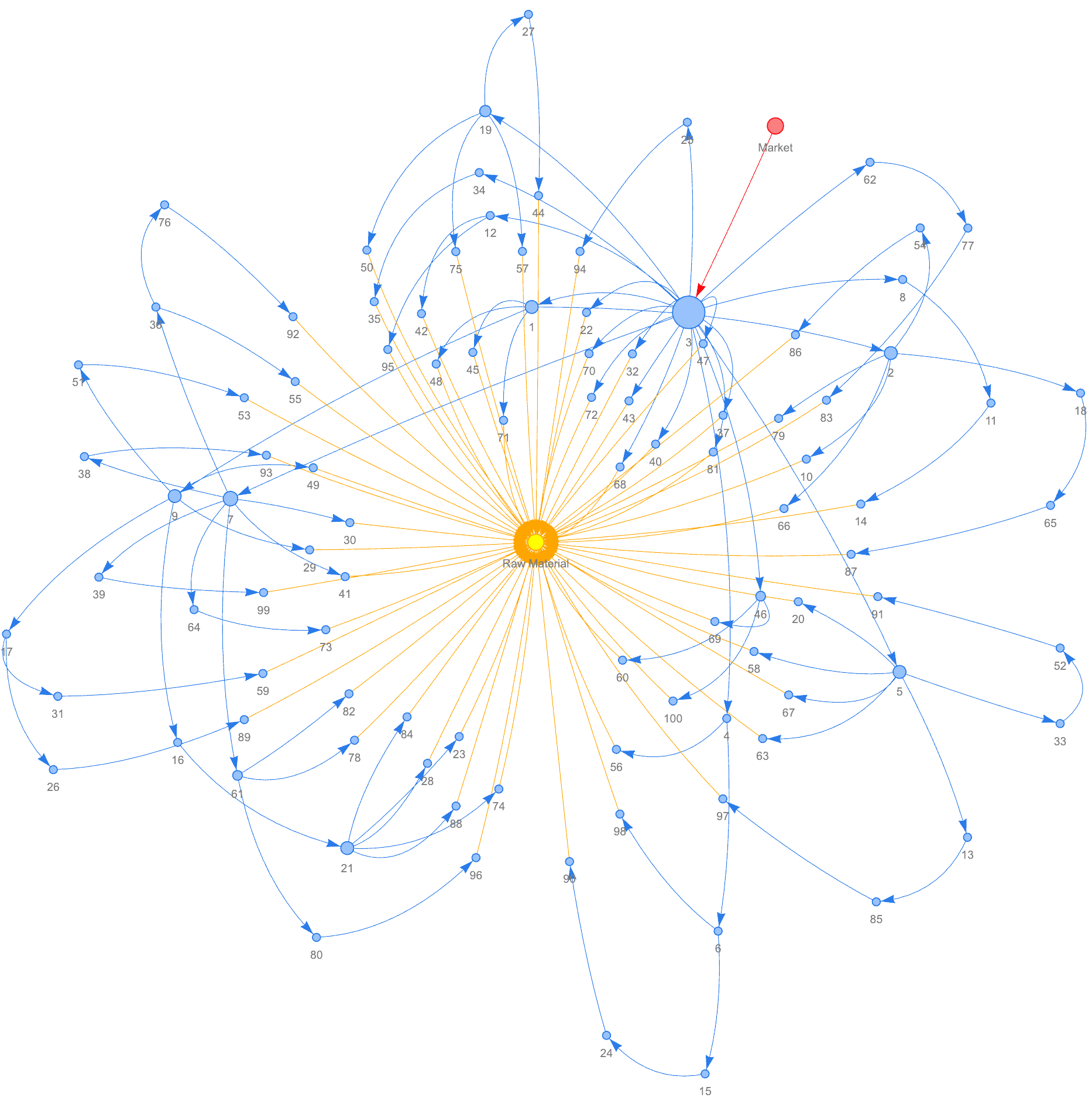}
    \caption{$\text{BA}(100,3), \delta = 3$ supply network: A visualisation of a scale-free, Barabasi-Albert, $\text{BA}(n,m)$ graph topology with structural modifications to represent a deep tier supply chain. $n=100$, $m = 3$, $\delta = 3$.}
    \label{fig:ba-graph}
\end{figure}

To generate random networks (Fig. \ref{fig:er_graph}), we use the Erdős–Rényi (ER) model, and for scale-free, structured networks (Fig. \ref{fig:ba-graph}), the Barabási–Albert (BA) model. These are classic generators of their respective topologies. Scale-free networks have consistent degree distribution across all scales and are structured. It has been suggested that supply networks typically have scale-free topology \cite{Brintrup2018SupplyField, Brintrup2016TopologicalIndustry, Brintrup2017SupplyCharacterization,Xu2023AutonomousLevels, Nair2011SupplyModel, Gafiychuk2000RemarksNetwork, Thadakamalla2004SurvivabilityPerspective}.

To capture the heterogeneous degree distribution of real supply chains, Fig. \ref{fig:ba-graph} compares the ER (random complex) and scale-free Barab\'{a}si-Albert (BA) models (structured complex) \cite{Barabasi1999EmergenceNetworks}. The BA model, $\text{BA}(N,m)$, is initialised with $m$ nodes. New nodes $u$ sequentially connect to $m$ existing nodes $v$, selecting them with probability proportional to $\text{deg}(v)$. Specifically, $p_{uv} = \frac{\text{deg}(v)}{\sum_{w \in V}\text{deg}(w)}$.

Both the ER and BA models produce undirected graphs, which we convert into directed acyclic graphs (DAGs) to define buyer-supplier relationships and cashflow direction. To create the DAG, we apply a unique total ordering on the nodes, based on their mean distance to all others, $l_u$. An undirected edge $(uv) \in E$ is made directed as $(uv)$ if $l_v > l_u$, or follows a preassigned random index if $l_v = l_u$. This preserves the total ordering and ensures a DAG structure \cite{Stanley1973AcyclicGraphs}.

We connect $m$ to all nodes with in-degree 0 and $r$ to all nodes with out-degree 0. This ensures focal nodes are those most connected to the network, while fringe nodes represent the producers of basic goods, as in automotive manufacturing networks. Node tiers $\mathcal{T}_u$ equal the directed geodesic length from the market node $m$ to node $u$ in the initial configuration. Starting cash positions are set as $\kappa_u^0 = 50 \mathcal{T}_u$.

Following \cite{Proselkov2024FinancialModel}, the power of nodes depends on the number of nodes in the same tier, $|\mathcal{T}_u|$. For the sets $S \ni s', M \ni m', L \ni l'$, for variables $\hat{s}, \hat{m}, \hat{l} \in [0,1]$, where $\hat{s} + \hat{m} + \hat{l} = 1$, we fix a proportion of the network to be small, medium, or large, such that $|S| = \hat{s}|V|, |M| = \hat{m}|V|, |L| = \hat{l}|V|$. Then, we create an ordering of all nodes $(u_1,...,u_n)$ where $|T_{u_1}| \leq ... \leq |T_{u_n}|$. We then fix power to nodes such that $P_u = l'$ for all $u \in \{u_1,...,u_{|L|}\}$, $P_v = m'$ for all $v \in \{u_{|L|+1},...,u_{|L|+|M|}\}$, and $P_w = s'$ for all $w \in \{u_{|L|+|M|+1},...,u_n\}$. This means that nodes in thinner tiers have greater power. Material prices are assigned iteratively along edges, starting from $r$. The price over edge $s_{ur}$ is set to 1. For antecedent edges, $s_{vu} = \max_{\{w:(uw) \in E\}}(s_{uw}) + 1$, ensuring profitability for all sales. Simulations were run on these DAGs.

\subsection{Distributed Collaborative Finance Threshold Control} \label{sec:extcomfinthr}
To inform the financial threshold function, we use nexus theory \cite{Yan2015APerspective}, which suggests that less embedded firms have less influence within the same supply chain. Embedding is defined from the firm's perspective and exists within its ego-network, the collection of firms it shares information with, bounded by visibility. Following \cite{Proselkov2022TheComputationb}, visibility is determined by geodesic distance. For a node $u \in V$, the set of nodes within $i$ edges is $H_i(u) \in G$, so a node $u$ with visibility $i$ shares information over $H_i(u)$.

Firms broadcast their financing policy $f_u^t$ to others within their shared ego-network. Node $u$ combines this data with its current expenses, $P_u^0$, weighting each value by a centrality measure relative to its ego-network. Centrality quantifies a node's importance in the network. To avoid over-sensitivity to current conditions, we apply a moving average with a time window $\hat{w}$. For a node $u$ and centrality measure $C(u; H_i(u))$, this is expressed as:
\begin{align*}
\hat{f_u^\tau} &= C(u; H_i(u)) P_u^0 + \sum_{v \in H_i(u)} C(v; H_i(u)) f_v^\tau\\
f_u^t &= \frac{\sum_{\tau = t-\hat{w}}^t \hat{f_u^\tau}}{\hat{w}}
\end{align*}
When the sum of centralities equals 0, we only use the information contained in $H_0$.

In our study, we examine four centrality measures: degree centrality ($C_d$), the total number of edges linked to a node; betweenness centrality ($C_b$), the ratio of paths through a node versus those bypassing it; eigencentrality ($C_e$), the node's influence as part of the eigenvector of the adjacency matrix; and closeness centrality ($C_c$), based on a node's average distance to others \cite{Freeman1977ABetweenness, Anderson1985EigenvaluesGraph}. These measures guide each node $u$ in determining $f_u^t$. We use all four to avoid bias toward any single benchmarking method.

By integrating policy broadcasting, centrality-based graph processing, financing threshold setting, and smoothing functions, we develop a distributed control mechanism for the supply network. This mechanism adjusts finance thresholds to mitigate ripple effect risk.

\subsection{Experimental and Analytical Procedure} \label{subsec:experiment}
We measure the survival time of supply network simulations using the nexus financing threshold model across the described topologies. Simulations are repeated over multiple configurations, varying graph topology, visibility, centrality measures, $\lambda$, and $e$. Each supply network has a diameter $\delta$, the shortest geodesic length in the graph, including market and raw materials nodes. Visibility is capped at $\delta$, as increasing it beyond this point does not change collaborative networks. To enable comparisons across graphs with varying $\delta$, we normalise visibility as relative visibility (RV) by dividing by $\delta$.

This produces histograms conditioned on RV, which we use to analyse the multimodality of system survival times by studying their probability distributions \cite{Proselkov2023FinancialModel}. Using kernel density estimates (KDEs) \cite{Rosenblatt1956RemarksFunction, Parzen1962OnMode}, we nonparametrically estimate these distributions. For the KDE bandwidth $h_D$ of a dataset $D$ from a simulation, we use a tuned version of Silverman's rule of thumb \cite{Silverman1986DensityAnalysis}, given by
\[
h_D = 0.9 \min\left(\sigma_D, \frac{\text{IQR}_D}{1.34}\right)|D|^{-\frac{1}{5}}\tilde{h},
\]
where for $D$, $\text{IQR}_D$ is interquartile range, $\sigma_D$ is standard deviation, and $|D|$ is dataset size. The tuning parameter, $\tilde{h}$, is constant across all $D$ for a given topology type.

To study how extended communication mitigates failure cascades, we apply breakpoint analysis using the binary segmentation algorithm \cite{Bai1997EstimatingTime} for a single breakpoint, $B_D$, on the KDE of $D$. The KDE before $B_D$, denoted $B_D^{\text{pre}}$, represents short-lived systems, while the KDE after $B_D$, denoted $B_D^{\text{post}}$, represents long-lived systems. We thus distinguish between short and long-lived system outcomes.

We analyse meaningful clusters by conducting modality analysis on $B_D^{\text{pre}}$ \cite{Proselkov2023FinancialModel}, counting prominent peaks. A peak $(x, y)$ is a local maximum with an adjacent local minimum $(x', y')$, where its prominence is $\bar{x} = y - y'$. A peak is prominent if $\bar{x} > \Xi$, with the prominence threshold $\Xi$ constant across all $D$. The number of prominent peaks in $D$ is denoted $\Lambda_D$.

To fit the KDEs for each $D$ and maximise the modality space, enhancing the precision of survivability analysis, we select $\tilde{h}$ to maximise $\Lambda_D$, constrained so $\min_D(\Lambda_D) = 1$. The value of $\tilde{h}$ is chosen to 5 decimal places of precision.

We then determine the set of long term survivors as all elements of the dataset with survival time greater than $B_D$, such that $D(B_D^{\text{post}})=\{t \in D: t\geq B_D\}$. Simulations are capped  at $t=1000$ if they reach it without breakage, and record this as their survival time. The LTVM is the proportion of survival times beyond the breakpoint. Given a space of simulations, $X$, it is defined as $\text{LTVM}^X = \frac{D(B_D^{\text{post}})}{D(0)}$. The IVM is restricted to the proportion of survival times at timestep $t=1000$. Given a space of simulations, $X$, it is defined as $\text{IVM}^X = \frac{D(1000)}{D(0)}$. We also investigate the relationship of RV to $\Lambda_D$, seeing how the complexity of the survival space changes as we vary visibility.

Polynomial regression is applied to each  output set, using the smallest polynomial degree with statistically significant $P$-value. We study their gradient if linear, and conduct detailed curve analysis of inflections and notable points.

\section{Results}
\label{sec:results}
Using our model, we investigated the relationship between visibility and viability in supply networks. Results are presented for varying RV$_D$ of the supply chain relative to diameter, with LTVMs, IVMs, and $\Lambda_D$ as output measures.

Experiments were conducted over the model parameter space listed in Table \ref{tab:ColVis}. Simulations ran on an Armari Magnetar workstation with a 3.5GHz Intel Xeon E5-1620 CPU, 64GB RAM, using Windows 10 Pro. Modelling and analysis were performed in Python 3 with the following packages: \texttt{os}, \texttt{pandas}, \texttt{tqdm}, \texttt{numpy}, \texttt{matplotlib}, \texttt{math}, \texttt{sklearn}, \texttt{itertools}, \texttt{pprint}, \texttt{art}, \texttt{networkx}, \texttt{random}, \texttt{sys}, \texttt{seaborn}, \texttt{statistics}, \texttt{time}, and \texttt{tabulate}.

\begin{table}[ht]
\footnotesize
\centering
\begin{tabular}{ll} 
 \hline
 Parameter & Values \\ [0.5ex]
 \hline
     Operation Fees $o$& 1\\
     Average Demand $\lambda$& 5, 10, 15\\
     Delivery Expenses $e$& 0.25, 0.75\\
     Loan Repayment Time& 120\\
     Firm Power Distribution& Heterogeneous\\
     Financed& True\\
     Paradigm& Proactive\\
     SCF Annual Rates $\gamma$& 20\%\\
     Bank Annual Interest Rates $\beta$& 10\%\\
     Invoice Term $\tau$& 60\\
     Window Size $\hat{w}$ & 60\\
     Medium Power Market Share $f(m)$& 3\\
     Large Power Market Share $f(l)$& 10\\
     Topologies& ER Random, BA Scale Free\\
     Network Diameters $\delta$& 4, 5, 6, 7\\
     Network Size $n$& 100\\
\hline
\end{tabular}
\caption{Table of the parameters tested.}
\label{tab:ColVis}
\end{table}

Generalised curves are generated to analyse relationships between output pairs and project behaviours to less formalised network topologies (Figs. \ref{fig:vis_kde_sf}, \ref{fig:sf-regs}, \ref{fig:vis_kde_rnd}, \ref{fig:rnd-regs}).  Our methodology preserves irreducible nonlinearities, studying the effects of varying visibility (and thus collaborative ranges) on long-term dynamics at finite and indefinite scales. Random and scale-free supply networks are tested as supply networks are often scale-free \cite{Brintrup2018SupplyField, Brintrup2016TopologicalIndustry, Xu2023AutonomousLevels, Ledwoch2018TheManagement}, with random topology serving as a null model.

\subsection{Scale Free Networks} \label{subsec:sf}
The scale free networks are parameterised such that $m = 1$ and $\delta \in \{4, 5, 6, 7\}$, representing a sparse supply chain of varying lengths.
\begin{figure}
    \centering
    \includegraphics[width=0.75\linewidth]{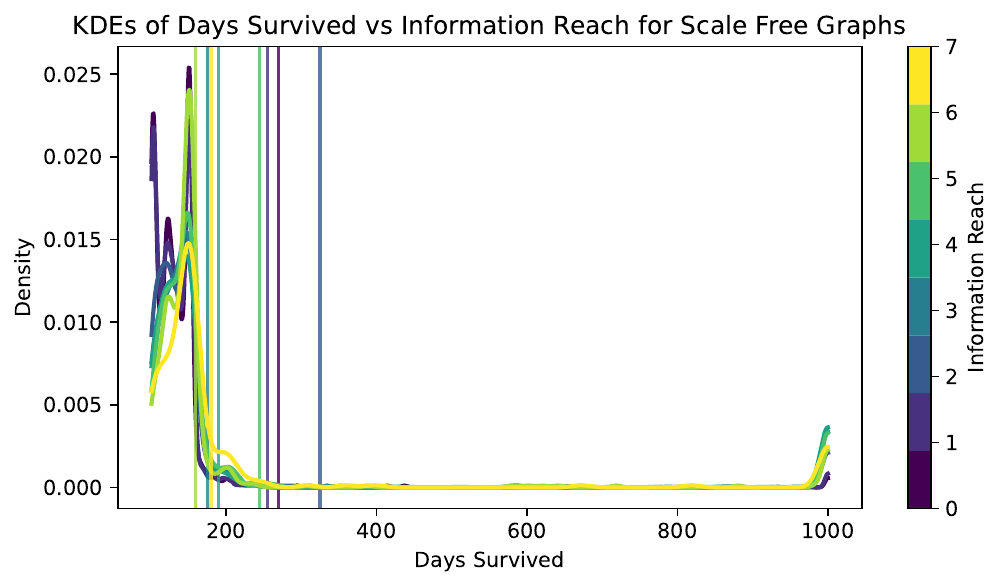}
    \caption{KDEs of Days Survived vs Information Reach for Scale Free Graphs: Vertical lines denote breakpoints separating short-lived supply chains and long-lived supply chains}
    \label{fig:vis_kde_sf}
\end{figure}

\begin{table}\centering
\scriptsize
\begin{tabular}{lrrrrr}\toprule
\textbf{$RV^\text{SF}_D$} & \textbf{$B_D^{\text{SF}}$} &\textbf{$\Lambda_D^\text{SF}$} &\textbf{IVM$_D^\text{SF}$} &\textbf{LTVM$_D^\text{SF}$} \\\midrule
0 &270 &11 &0.0042 &0.0115 \\
0.1429 &230 &5 &0.0125 &0.0208 \\
0.1667 &280 &16 &0 &0.0042 \\
0.2 &200 &3 &0.0083 &0.0083 \\
0.25 &195 &3 &0.0125 &0.0125 \\
0.2857 &315 &5 &0.0333 &0.05 \\
0.3333 &375 &17 &0 &0 \\
0.4 &340 &4 &0.0458 &0.05 \\
0.4286 &150 &6 &0.0583 &0.5208 \\
0.5 &225 &8 &0.0563 &0.0854 \\
0.5714 &375 &3 &0.075 &0.0833 \\
0.6 &190 &4 &0.0958 &0.1625 \\
0.6667 &190 &6 &0.0167 &0.0458 \\
0.7143 &260 &3 &0.0917 &0.1042 \\
0.75 &335 &1 &0.1333 &0.1458 \\
0.8 &205 &2 &0.1208 &0.1542 \\
0.8333 &170 &6 &0.0125 &0.0667 \\
0.8571 &160 &8 &0.0667 &0.2292 \\
1 &180 &2 &0.0875 &0.1469 \\
\bottomrule
\end{tabular}
\caption{All outputs investigated for the scale-free topologies}\label{tab:sf_outputs}
\end{table}

Figure \ref{fig:vis_kde_sf} shows the KDEs of survival time distributions conditioned on visibility for scale-free graphs. Vertical lines indicate $B_D$ for each curve, colour-coded accordingly. Visibility curves are used instead of RV for clarity.

Table \ref{tab:sf_outputs} presents the outputs of the investigated measures, conditioned on RV$_D$. Polynomial regression curves for relevant pairs are shown in Fig. \ref{fig:sf-regs}, with detailed descriptions in the appendix.

\begin{figure}
    \centering
    \begin{subfigure}{0.49\textwidth}
        \includegraphics[width=\textwidth]{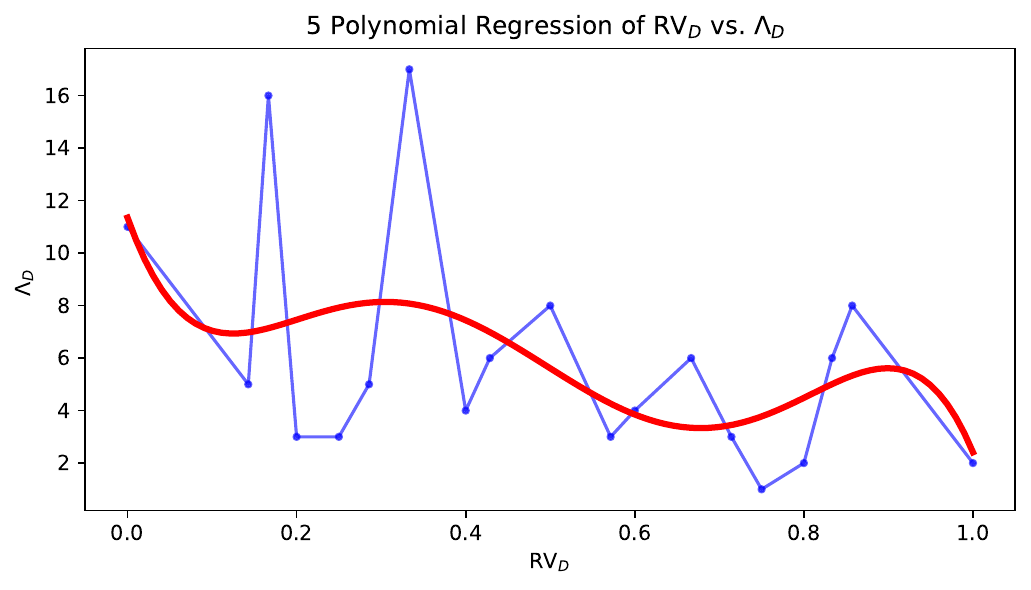}
        \caption{RV$_D$ vs $\Lambda^\text{SF}_D$ with a $5^{\text{th}}$ degree polynomial regression.}
        \label{fig:rv-lambda-sf}
    \end{subfigure}
    \begin{subfigure}{0.49\textwidth}
        \includegraphics[width=\textwidth]{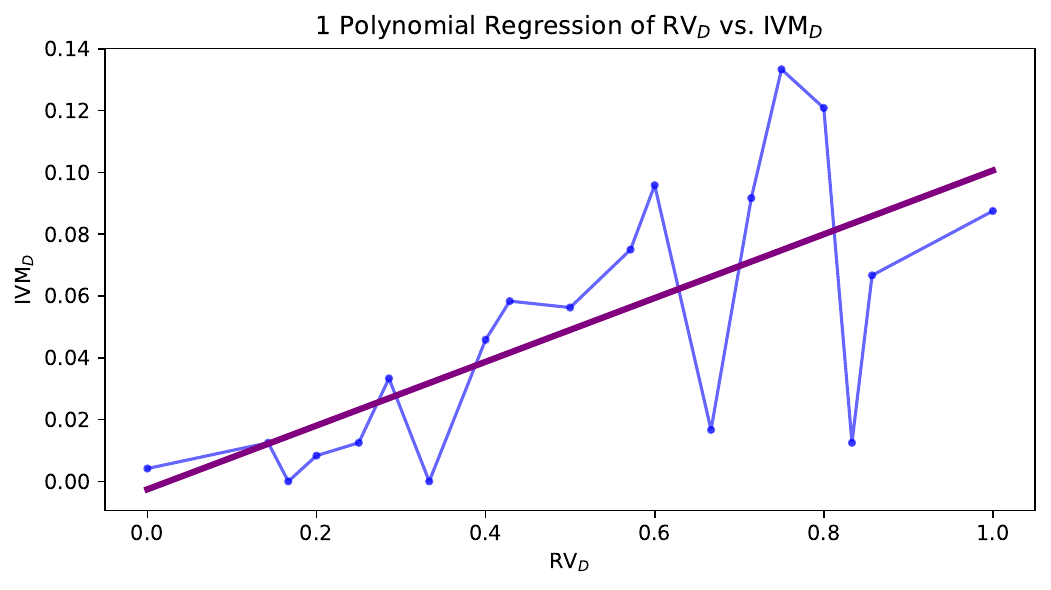}
        \caption{$\text{RV}_D$ vs $\text{IVM}^\text{SF}_D$ with a $1^{\text{st}}$ degree polynomial regression.}
        \label{fig:rv-ivm-sf}
    \end{subfigure}
    \begin{subfigure}{0.49\textwidth}
        \includegraphics[width=\textwidth]{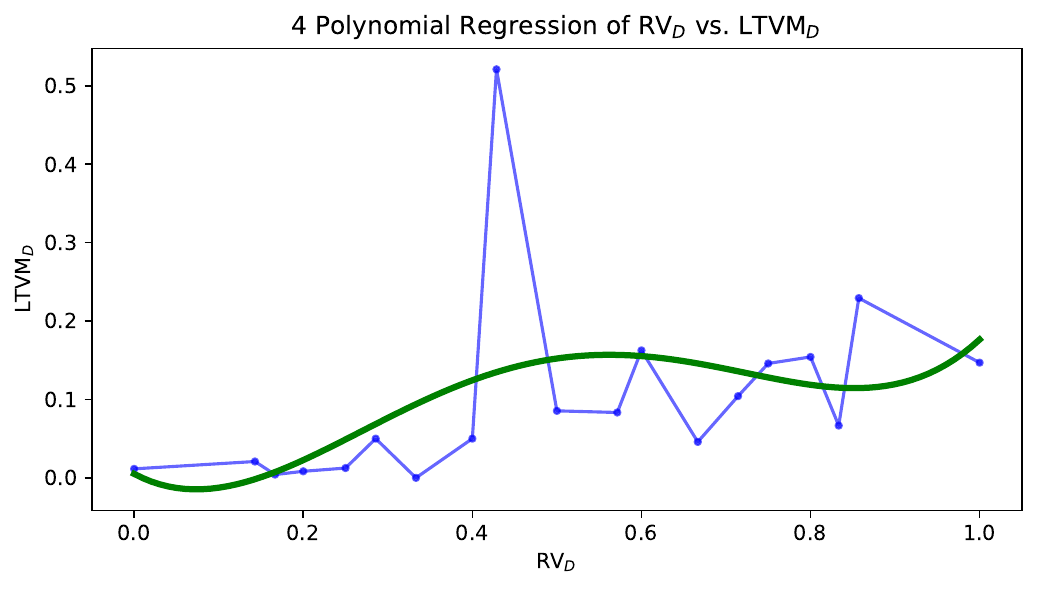}
        \caption{$\text{RV}_D$ vs $\text{LTVM}^\text{SF}_D$ with a $4^{\text{th}}$ degree polynomial regression. Prominent value at $(0.4286, 0.5208)$.}
        \label{fig:rv-ltvm-sf}
    \end{subfigure}
    \begin{subfigure}{0.49\textwidth}
        \includegraphics[width=\textwidth]{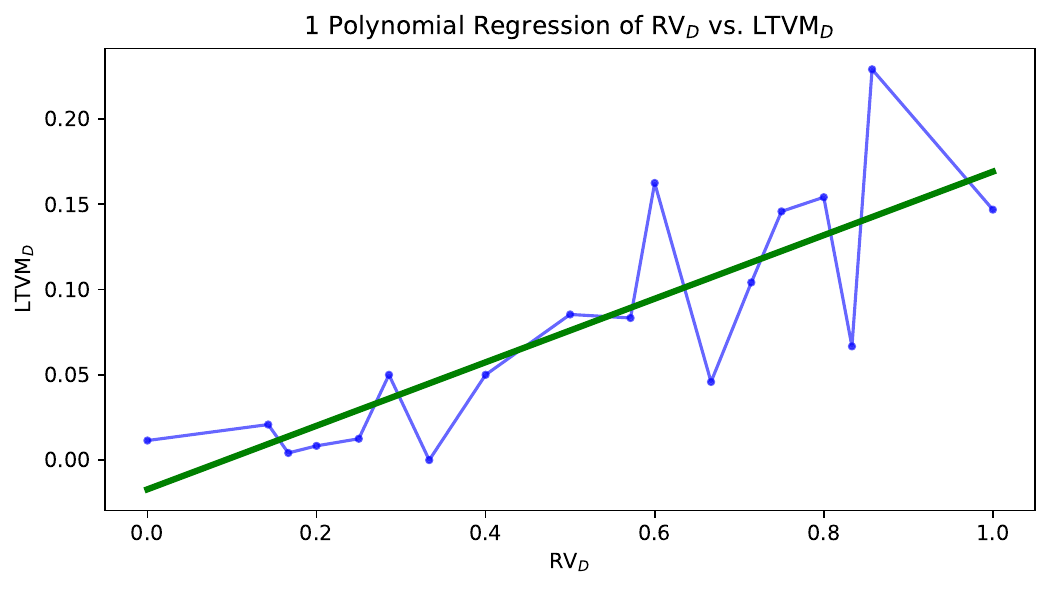}
        \caption{$\text{RV}_D$ versus $\text{LTVM}^\text{SF}_D$, omitting $(0.4286, 0.5208)$ with a $1^{\text{st}}$ degree polynomial regression.}
        \label{fig:rv/rv0.4286-ltvm-sf}
    \end{subfigure}
    \centering
    \caption{Regressions of $\text{RV}_D$ to $\Lambda^\text{SF}_D$, $\text{IVM}^\text{SF}_D$, and $\text{LTVM}^\text{SF}_D$ for scale-free networks}
    \label{fig:sf-regs}
\end{figure}

For scale-free topologies, there is a high-density region before $B_D$, followed by a long plateau with noise and a prominent inflection near $t=1000$ (Fig. \ref{fig:vis_kde_sf}). Therefore, imparting a local collaboration function with variable visibility to all supply network members enhances the network's viability. 

Nonlinearities in the scale-free topology curves suggest target viabilities may be achieved with smaller, more attainable visibilities. $\text{IVM}^\text{S}_D$ reaches nearly 14\% (Fig. \ref{fig:rv-ivm-sf}), while $\text{LTVM}^\text{S}_D$ exceeds 52\% (Fig. \ref{fig:rv-ltvm-sf}), or 22\% excluding a prominent value (Fig. \ref{fig:rv/rv0.4286-ltvm-sf}). Optimal solutions depend on which viability is prioritised, as the globally maximal $\text{RV}_D$ and local extrema differ between $\text{LTVM}^\text{S}_D$ and $\text{IVM}^\text{S}_D$.

The regression for peak count $\Lambda^\text{SF}_D$ in scale-free topologies shows a negative slope with $\text{RV}_D$ (Fig. \ref{fig:rv-lambda-sf}). Peaks in $\Lambda_D^\text{SF}$ represent local maximum likelihoods of survival time caused by clustering, indicating lower variability. As $\text{RV}_D$ increases, variability in survival times grows, reducing clustering in short-lived networks and improving viability.

\subsection{Random Networks} \label{sebsec:rnd-net}
Next, we examine the random networks. These were parameterised as $(p, \delta) \in \{(0.1,7),$ $(0.2,6), (0.2,5), (0.3,4)\}$. This setup explores a range of diameters, and since diameter is a function of density, distinct connection probabilities were used for different diameters.
\begin{figure}
    \centering
    \includegraphics[width=0.75\linewidth]{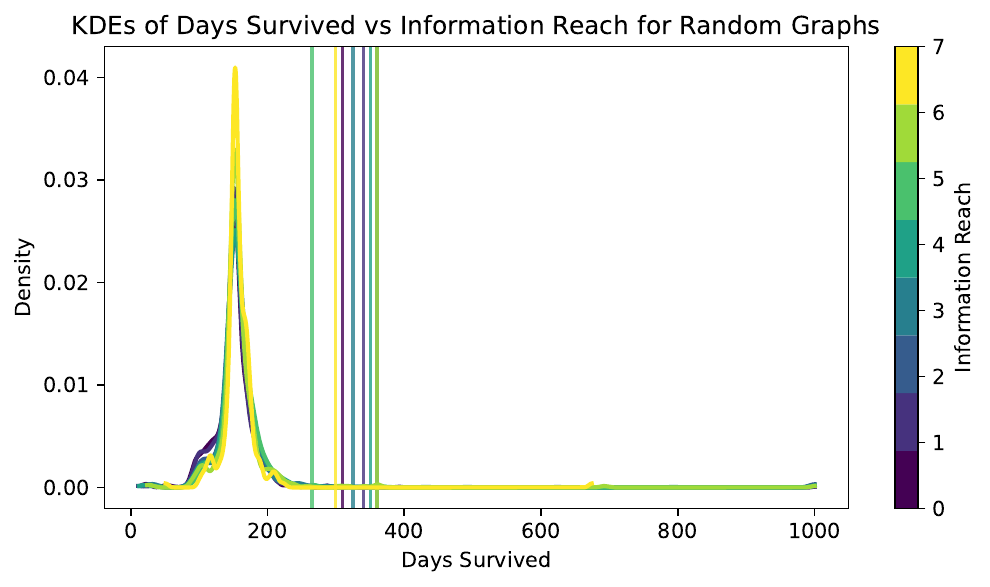}
    \caption{KDEs of Days Survived vs Information Reach for Random Graphs: Vertical lines denote breakpoints separating short-lived supply chains and long-lived supply chains}
    \label{fig:vis_kde_rnd}
\end{figure}

\begin{table}[!htp]\centering
    \scriptsize
    \begin{tabular}{lrrrrr}\toprule
        \textbf{$RV^\text{R}_D$} & \textbf{$B_D^{\text{R}}$} &\textbf{$\Lambda_D^\text{R}$} &\textbf{IVM$_D^\text{R}$} &\textbf{LTVM$_D^\text{R}$} \\\midrule
        0 &310 &10 &0.001042 &0.003125 \\
        0.142857 &570 &18 &0 &0 \\
        0.166667 &270 &5 &0.004167 &0.016667 \\
        0.2 &540 &9 &0 &0 \\
        0.25 &740 &4 &0 &0 \\
        0.285714 &395 &21 &0 &0 \\
        0.333333 &485 &6 &0 &0 \\
        0.4 &370 &3 &0.008333 &0.008333 \\
        0.428571 &775 &16 &0 &0 \\
        0.5 &355 &1 &0.0125 &0.016667 \\
        0.571429 &430 &15 &0 &0 \\
        0.6 &315 &4 &0.004167 &0.0125 \\
        0.666667 &650 &5 &0 &0 \\
        0.714286 &220 &5 &0.004167 &0.016667 \\
        0.75 &290 &5 &0.004167 &0.0125 \\
        0.8 &255 &4 &0.004167 &0.0125 \\
        0.833333 &565 &6 &0 &0 \\
        0.857143 &225 &10 &0 &0.008333 \\
        1 &365 &5 &0.004167 &0.005208 \\
        \bottomrule
    \end{tabular}
    \caption{All outputs investigated for the random topologies}\label{tab:rnd_outputs}
\end{table}

Fig. \ref{fig:vis_kde_rnd} shows the KDEs of survival time distributions conditioned on visibility for random graphs. To assess minimal remaining viability, we compare viable network distributions (Table \ref{tab:rnd_outputs}), visualised in Fig. \ref{fig:rnd-regs}. Further details are in the appendix.

\begin{figure}
    \centering
    \begin{subfigure}{0.48\textwidth}
        \includegraphics[width=\textwidth]{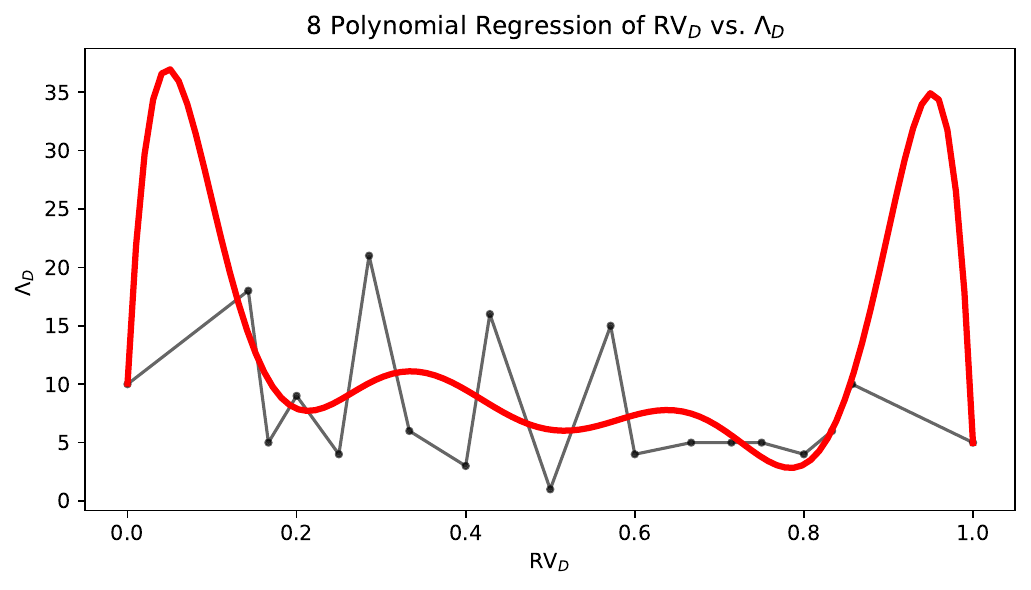}
        \caption{RV$_D$ vs $\Lambda^\text{R}_D$ with an $8^{\text{th}}$ degree polynomial regression.}
        \label{fig:rv-lambda-rnd}
    \end{subfigure}
    \begin{subfigure}{0.49\textwidth}
        \includegraphics[width=\textwidth]{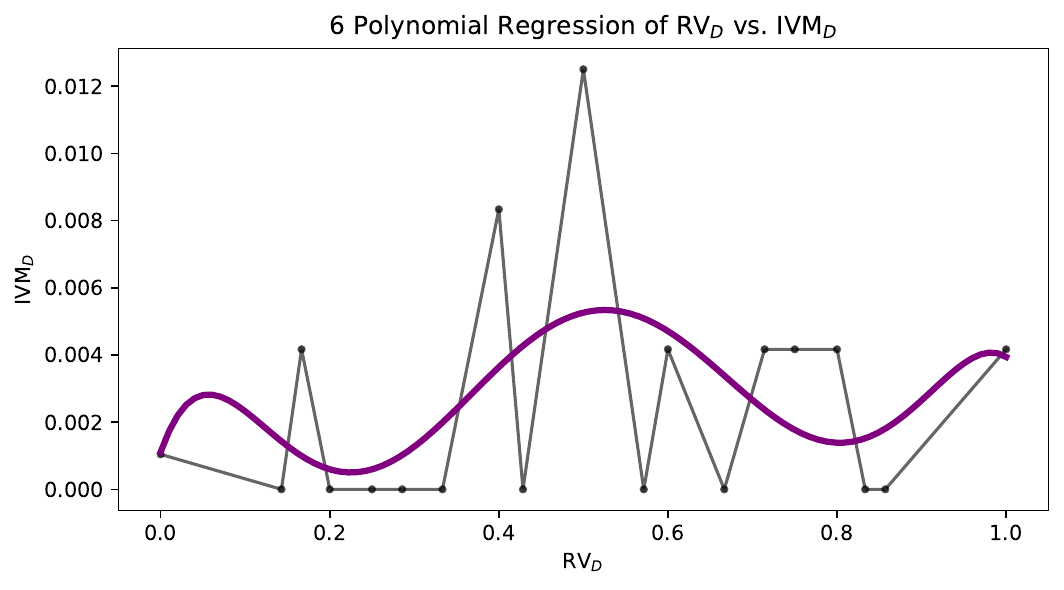}
        \caption{$\text{RV}_D$ versus $\text{IVM}^\text{R}_D$ with a $6^{th}$ degree polynomial regression}
        \label{fig:RVD_D1000}
    \end{subfigure}
    \begin{subfigure}{0.49\textwidth}
        \includegraphics[width=\textwidth]{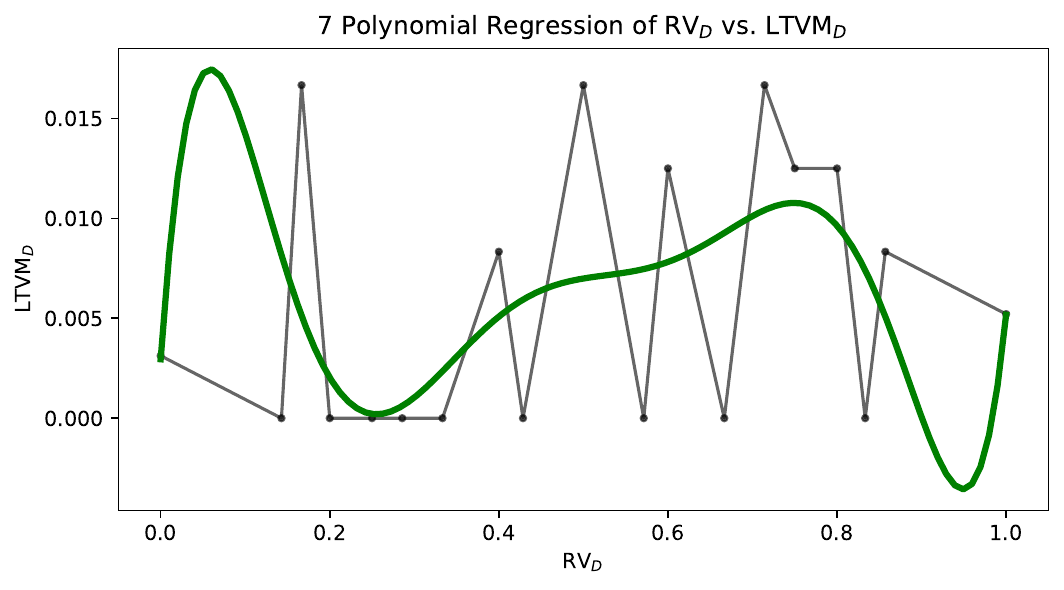}
        \caption{$\text{RV}_D$ vs $\text{LTVM}^\text{R}_D$ with a $7^{\text{th}}$ degree polynomial regression}
        \label{fig:RV_LTVM}
    \end{subfigure}
    \caption{Regressions of $\text{RV}_D$ to $\Lambda^\text{SF}_D$, $\text{IVM}^\text{SF}_D$, and $\text{LTVM}^\text{SF}_D$ for random networks}
    \label{fig:rnd-regs}
\end{figure}

Short-lived networks exhibit strong clustering around a single point, indicating a bottleneck, with flatter, more ordered curves compared to scale-free graphs (Fig. \ref{fig:vis_kde_sf}) and an order of magnitude smaller peaks at the KDE ends. Random networks are thus most likely to fail at a single point, regardless of visibility, with near-zero likelihood of viability.

The maximum point for the random data is $(0.5, 0.0125)$, giving $\max \text{IVM}^\text{R}_D = 0.0125$. We have that, $\frac{\max \text{IVM}^\text{SF}_D}{\max \text{IVM}^\text{R}_D} = 10.664$, indicating that scale-free structure significantly enhances the impact of collaboration (Fig. \ref{tab:rnd_outputs}). In random graphs, relevancy principles break down, suggesting relevancy depends on meaningful graph structure. Other topologies with scale-free-like properties, such as hierarchy or scaling, may also support viability.

In scale-free topologies (Fig. \ref{fig:sf-regs}), long-term and indefinite viability occur with similar frequency per RV$_D$, accounting for potential outliers. Viability positively correlates with visibility, as shown by the regression slopes in Fig. \ref{fig:rv-ivm-sf}, Fig. \ref{fig:rv-ltvm-sf}, and Fig. \ref{fig:rv/rv0.4286-ltvm-sf}. In random networks, viability shows a nonlinear relationship with visibility, with substantial variation around the curves. Noise causing this is unlikely since the dataset spans multiple configurations and topologies, explaining away noisiness. This shows a nonlinear relationship between visibility and viability. Notably, $\text{RV}_D = 1$ does not maximise viability, suggesting that centralised control is unscalable with network size, while distributed control with limited collaboration  yields superior viability of all types.

\section{Discussion} \label{sec:disc}
In this study, we explored the creation of a scalable, autonomously controlled supply network, mitigating ripple effect risk by providing firms with information about a limited number of subjectively relevant peers. Using a complex network agent-based simulation model, we captured supply network dynamics to test our hypothesis: limited, up-to-date visibility enables scalable network autonomy.

To mitigate against ripple effect risk with a feasible privacy-preserving function, firms communicated financial policies to inform distributed cash control, achieved via real-time broadcasting among local, limited regions (ego-networks) around given agents. 
We found that this resulted in a scalable network that autonomously controlled ripple effects, achieving systemic viability. We characterised the relationship between visibility and viability, showing how ego-network size influences dynamic adaptivity and recovery from shocks \cite{Ivanov2020ViablePandemic}.

Ripple effect risk in our simulation was endogenous and systemic, rather than exogenous and shock-derived, necessitating holistic views enabled by visibility \cite{Bimpikis2018MultisourcingNetworks}. While isolating steady states and risk sources is challenging, our simulation allowed us to bypass this problem and measure network performance holistically. We introduced novel viability measures to capture long-term performance, using these in our analysis.

Our hypothesis was validated, and in constructing our model, we defined a risk mitigation approach using collaborative intelligence \cite{Devadasan2013CollaborativePlanning}. We developed a bidirectional risk mitigation function and tunable form of distributed autonomous control via SCF, contributing a method that advances supply networks toward true autonomy \cite{Xu2024TowardsLevels, Xu2023AutonomousLevels}.

We introduced a novel control method to maximise viability in a deep-tier interfirm upstream supply network. Using limited visibility and nexus theory-driven collaboration, this approach autonomously generated financing thresholds to adjust cash positions via SCF, ensuring optimal liquidity across the network.

Since viability was attained with only limited information for each firm in the supply network, this suggests data collection expenses and efforts for firms  need not expand at the same rate as supply network growth. From this, we conclude that we have defined scalable autonomy for our risk case. Achieving viability with limited information per firm suggests that data collection efforts need not scale with network growth, defining scalable autonomy for our risk case. Moreover, the absence of a supervisory control agent highlights this as a form of distributed, locally decentralised control. This suggests that such control enables scalable autonomy, demonstrating the potential of distributed cooperation in production economies.

Thresholds were defined to preserve privacy, as policy information depended on partner firms through a local centrality function. This allowed risk information to propagate quickly across the network, likely facilitating mitigation. It may thus be that interfirm privacy not only supports but enables effective risk mitigation. Further investigation into the relationship between visibility and the propagation rates of risk and information could confirm this.

Outputs for scale-free networks suggest visibility impacts viability in phases, allowing certain visibility levels to optimise viability more efficiently. We propose this comes from the synchronisation of shock and information propagation in the control response. Further research into network-theoretical representations of these effects could enable true viability optimisation, offering new opportunities for business network management \cite{Dekker2022HiddenComplexity}.

The positive relationship between visibility and viability highlights the importance of collaboration, while nonlinear phases highlight the importance of focusing on relevant partners. This aligns with inventory management research \cite{Zhu2024EffectsSimulation} and proactive financing studies \cite{Ivanov2024CashDisruptions}, suggesting broader applicability for operational risk mitigation.

Our test environments were complex deep-tier supply network topologies, derived from theoretical complex networks, modified to represent acyclic firm-to-firm upstream supply networks. We introduced a testing methodology that analyses dynamic network data using novel simulation-based viability metrics. These metrics measured viability as the proportion of simulations surviving past a data-derived threshold for finite viability (LTVM) or a preselected simulation termination threshold for indefinite viability (IVM). LTVM and IVM are distinguished because survival times at $t=1000$ include legitimate failures and firms cut off by capped simulation times. Practically, long-term viability (LTVM) suits temporary measures, while indefinite viability (IVM) is preferable for permanent institutional chains.

The simulated environment shows multimodality in survival distributions \cite{Proselkov2024FinancialModel}, prompting the use of a data-derived threshold to separate long and short term survival spaces. Simulation survival time distributions were found by generating kernel density estimates (KDEs), with bandwidths chosen to maximise cluster quantity spread in short-term modes while ensuring at least one distribution had only one cluster. We analysed the distribution of peaks and the survivability threshold, revealing that greater visibility reduces the number of clusters but shows no linear relationship with the threshold (Fig. \ref{fig:rv-lambda-sf}, Fig. \ref{fig:rv-lambda-rnd}).

Competition within a supply chain prunes suboptimal configurations, enabling survival, improvement, and quality delivery \cite{Gofman2020ProductionDestruction}. While visibility does not extend the lifespan of short-lived supply networks, it increases the variety of failures by reducing peak counts, suggesting that visibility preserves and enhances the benefits of competition.  These findings suggest that distributed cooperation, when implemented holistically, has minimal downsides. Further research should explore the relationship between visibility and competitiveness in greater detail, focusing on asymmetric visibility and conducting rich statistical analyses to assess the impact of reduced collaboration forms.

Note the ``outlier'' in Fig.~\ref{fig:rv-ltvm-sf}. Excluding it in Fig.~\ref{fig:rv/rv0.4286-ltvm-sf} reveals a linear visibility-viability relationship. However, since each datapoint aggregates multiple simulations, treating it as an outlier may be inappropriate. With this in mind, we demonstrated that limited visibility can greatly outperform complete visibility. Both conclusions are true under different conditions.

These are simulated environments. In empirical research or supply chain management, statistical significance is used to infer whether an observed effect is likely due to chance or represents a real phenomenon in the population \cite{Pfeffer2006Evidence-basedManagement}. When modelling entire systems without sampling variability, the concept of statistical significance may not be directly applicable or necessary \cite{Gelman2006TheSignificant}, as meaning exists without it \cite{Cohen1994The05}. Unlike supply chain managers, who face information limits, simulations reveal entire-system dynamics, allowing valid conclusions without reliance on statistical significance \cite{Winsberg2010ScienceSimulation}.

The discrepancy between our regression analysis and results suggests two paradigms: statistical and simulated. The statistical paradigm assumes significance in incomplete data scenarios, implying that it may be appropriate for policymakers ignorant of the evolution scenarios of the supply network to assume that viability increases with visibility \cite{Caridi2014TheModel}. Conversely, under the simulation paradigm, which models entire system behaviour, we conclude that maximising visibility is suboptimal due to peaks and valleys in the output dataset. Despite differences, both paradigms can inform each other \cite{Morgan2005ExperimentsSurprise}.

For policymakers, confidence in the simulation paradigm requires a thorough understanding of supply network evolution scenarios. We recommend a subsumptive strategy: increase visibility while improving scenario understanding, given their risk mitigation strategy. Once sufficient confidence is achieved, select a visibility level that meets multivariate supply chain risk management targets, including viability. The exceptional performance of specific configurations (Fig. \ref{fig:sf-regs}, Fig. \ref{fig:rnd-regs}) highlights the potential for precision engineering of supply network topologies to maximise viability, leveraging the research environment developed in this study.

None of the regressions for random topologies show a general incline, though IVM$_D^\text{R}$ performs best at RV$_D = 0.5$ (Fig. \ref{fig:rnd-regs}). While ER graphs have no inherent scaling or hierarchy, converting them into DAGs introduces structure due to the market and raw materials nodes, $m$ and $r$. The observed association between visibility and indefinite viability in random graphs may result from most nodes having both $m$ and $r$ visible at RV$_D = 0.5$. Further investigation is needed to test scenarios where nodes lack visibility into $m$ or $r$.

Altogether, results indicate that scale-free topology significantly enhances viability, increasing IVM by an order of magnitude and LTVM by two orders compared to random topologies. In contrast, random topologies severely restrict viability.  In both topologies, visibility and viability exhibit a nonlinear relationship, however, in scale-free topologies, viability of both sorts tends to grow with visibility.

In scale-free supply networks, increased visibility reduces bottlenecks, leading to fewer short-lived survival clusters (Fig. \ref{fig:vis_kde_sf}) and broader outcome ranges, enhancing viability. Our findings show a negative association between clustering in unviable spaces and viable network size, with less clustering reducing dependency on key risks. This variety, influenced by visibility, empowers firms by enabling more diverse responses to shocks, improving long-term survival. To maximise viability, supply-chain managers should promote dynamic network configurations and include key players in policy decisions, reflecting the interconnectedness of global financial risk \cite{Leibrock2022InterconnectednessRevisited, chen2023TheEvidence}. However, the post-COVID trend toward regionalised supply chains \cite{Barthe-Dejean2021ShiftingMatrix, Goldberg2023IsDeglobalizing} may hinder viability by limiting coverage and increasing financial risk propagation, so we recommend reintegration and collaborative regional policies to counter this.

This study extends on previous research \cite{Proselkov2023FinancialModel}, where financial thresholds were based on a firm's historic behaviour. We instead use a function of the present behaviour of other firms. The previous study showed that nexus methods can outperform temporal methods for risk mitigation, and we build on this to demonstrate viability. This argues in favour of peer-to-peer communication or transaction platforms \cite{Wei2016P2PModel, Pan2021P2PEfficiency}, rather than optimising visibility for the focal firm. Future research should combine temporal and nexus information \cite{Randall2009SupplyChain, Hofmann2007TheChain, Hofmann2011StrengtheningOptimization, Hofmann2021SupplyHealthy}.

The methods outlined in this paper provide a foundation for further research. Applied to an empirical supply chain and tuned to its dynamics, the simulation could enable meaningful perturbation analysis to evaluate supply chain quality. This could guide the design of limited visibility information-sharing schemes to enhance financial viability within specific supply ecosystems.

\section{Acknowledgements}
This research was supported by the both EPSRC and BT Prosperity Partnership project: Next Generation Converged Digital Infrastructure, grant number EP/R004935/1, and the UK Engineering and Physical Sciences Research Council (EPSRC) Doctoral Training Partnership Award for the University of Cambridge, grant number EP/R513180/1.

\section{Data Availability Statement}
The data that support the findings of this study are available from the corresponding author, Y. P., upon reasonable request.

\appendix

\section{Appendix}
\label{sec:appendix}
The following section details our analysis and description of the polynomial regression curves of the relationships of various parameters in our analysis.

\subsection{Polynomial Regression Curves}
This subsection describes polynomial regression curves of relevant pairs inside  Table \ref{tab:sf_outputs}, visualised by Fig. \ref{fig:sf-regs}, to understand all elements further.

\subsubsection{Scale-Free Diagrams}
\paragraph{RV$_D$ vs $\Lambda^\text{SF}_D$}
First,  for the scale-free diagrams,  we study the relationship between relative visibility, RV$_D$ and the number of peaks before the breakpoint, $\Lambda_D$. This has a $5^{\text{th}}$-degree polynomial, accurate with $P = 0.027$ and standard error $\sigma_{\bar{D}} = 0.414$. The regression curve (Fig. \ref{fig:rv-lambda-sf}) is defined according to:
\begin{align*}
\widehat{\Lambda^{\text{SF}}_D}(\text{RV}_D) = 
&-780.0891 \text{RV}_D^5 + 1956.8451 \text{RV}_D^4\\
&-1723.1095 \text{RV}_D^3 + 627.9353 \text{RV}_D^2\\
&-90.5196 \text{RV}_D + 11.3497.
\end{align*}

\paragraph{$\text{RV}_D$ vs $\text{IVM}^\text{SF}_D$}
For the scale free graphs, measuring relative visibility, $\text{RV}_D$, to proportion of all failures at $t=1000$, $\text{IVM}^\text{SF}_D$, we have a $1^\text{st}$ degree polynomial regression, accurate with $P=0.001$ but a standard error, $\sigma_{\bar{D}} = 0.255$. The regression curve (Fig. \ref{fig:rv-ivm-sf} is defined according to
$$
\widehat{\text{IVM}^{\text{SF}}}(\text{RV}_D) = 0.1032 \text{RV}_D - 0.0026,
$$
where $\widehat{\text{IVM}^\text{SF}_D(1)} = 0.1006$.

\paragraph{$\text{RV}_D$ vs $\text{LTVM}^\text{SF}_D$} 
For the scale-free graph, measuring the proportion of values past the break point, $\text{LTVM}^\text{SF}_D$, we have significance, with $P = 0.041$ and $\sigma_{\hat{D}} = 0.451$, for a $4^{\text{th}}$ order polynomial regression that. The curve (Fig. \ref{fig:rv-ltvm-sf}) is defined as 
$$
\widehat{\text{LTVM}^\text{SF}_D}(\text{RV}_D) = 0.0051 - 0.5797 \text{RV}_D + 4.8083 \text{RV}_D^2 - 8.1847 \text{RV}_D^3 + 4.1274 \text{RV}_D^4.
$$

\paragraph{$(\text{RV}_D; \text{RV}_D \neq 0.4286)$ vs $\text{LTVM}^\text{SF}_D$}
The regression Fig. (\ref{fig:rv/rv0.4286-ltvm-sf}) is defined according to
$$
\widehat{\text{LTVM}^\text{SF}_D}(\text{RV}_D; \text{RV}_D \neq 0.4286) = -0.0172 + 0.1863\text{RV}_D,
$$ 
and $P = 0$ and $\sigma_{\hat{D}} = 0.017$. 

\subsubsection{Random Networks}
The following contains polynomial regression analysis of relevant pairs inside this table, to understand all elements further.

\paragraph{$\text{RV}_D$ vs $\Lambda^\text{R}_D$}
We repeat our analysis for the random graphs, comparing $\Lambda^\text{R}_D$ to $\text{RV}_D$. The polynomial regression (Fig. \ref{fig:rv-lambda-rnd}) delivers an $8^{\text{th}}$ degree polynomial, accurate with $P = 0.031$, with $\sigma_{\bar{D}} = 0.425$. It is defined as 
\begin{align*}
\widehat{\Lambda^\text{R}_D}(&\text{RV}_D) = \\
& - 207895.4284\text{RV}_D^8 + 828388.3627\text{RV}_D^7\\
& - 1358506.6894\text{RV}_D^6 + 1182795.9815\text{RV}_D^5\\
& - 586913.5003\text{RV}_D^4 + 164780.6428\text{RV}_D^3\\
& - 24049.5475\text{RV}_D^2 + 1395.0947\text{RV}_D\\
& + 10.0570.
\end{align*}

\paragraph{$\text{RV}_D$ vs $\text{IVM}^\text{R}_D$}
We measure relative visibility, $\text{RV}_D$, to proportion of all failures at $t=1000$, $\text{IVM}^\text{R}_D$, we have a $6^\text{th}$ degree polynomial regression (Fig. \ref{fig:RVD_D1000}), accurate with $P=0.05$ exactly, but a standard error, $\sigma_{\bar{D}} = 0.473$. The curve is defined according to 
\begin{align*}
\widehat{\text{IVM}^\text{R}_D}(\text{RV}_D) = 
& - 2.2411\text{RV}_D^6 + 6.9729\text{RV}_D^5\\
& - 8.0382\text{RV}_D^4 + 4.1713\text{RV}_D^3\\ 
& - 0.9332\text{RV}_D^2 + 0.0711\text{RV}_D\\
& + 0.0011.
\end{align*}

\paragraph{$\text{RV}_D$ vs $\text{LTVM}^\text{R}_D$}
Finally we cover $\text{RV}_D$ vs $\text{LTVM}^\text{R}_D$. We have a $7^\text{th}$ degree polynomial (Fig. \ref{fig:RV_LTVM}) in this case, with $P = 0.040$, and $\sigma_{\bar{D}} = 0.449$, defined by:
\begin{align*}
    \widehat{\text{LTVM}^\text{R}_D}(\text{RV}_D) = 
&28.7872\text{RV}_D^7 -102.8713\text{RV}_D^6\\
&+ 147.1856\text{RV}_D^5 -107.4140\text{RV}_D^4\\
&+ 41.8254\text{RV}_D^3 -8.1099\text{RV}_D^2\\
&+ 0.5993\text{RV}_D + 0.0029.
\end{align*}

\subsection{Comparative Regression Analysis}
To account for the nonlinearity in the regressions, we subdivide sections into distinct regions, specifically the low ($L$), mid-low (ML), mid ($M$), mid-high (MH), and high ($H$) regions, defined for $d \in D$ as,
\begin{align*}
    L &:= [0,0.1667]\\
    \text{ML} &:= (0.1667, 0.3333]\\ 
    M &:=(0.3333, 0.6667]\\
    \text{MH} &:=(0.6667, 0.8333]\\
    H &:=(0.8333, 1].
\end{align*}

\subsubsection{Scale Free Networks}
In Fig. \ref{fig:vis_kde_sf}, which shows the KDEs for each distribution of survival times conditioned on visibility for the scale-free graphs, we see that short-lived supply networks are most common, with a mean of 0.8999 simulations of a given RV terminating before $B_D$. After this, a mean of 0.0511 simulations of a given RV are long-term but finite, occupying the long, flat, near zero region after each $B_D$. Finally, a mean of 0.0490  of a given RV last indefinitely, inducing the peak at the end of the KDEs.

\paragraph{RV$_D$ vs $\Lambda^\text{SF}_D$}
The regression (Fig. \ref{fig:rv-lambda-sf}), with clear downward incline, has standard error $\sigma_{\bar{D}} = 0.414$, likely due to a triangular pattern in the data, as well as the exceptionally high points of $(0.2857, 21)$ and $(0.1429, 18)$, affecting the similarity to smooth approximations. Since
$\widehat{\Lambda^{\text{SF}}_D}(0) = 11.3497$ and $\widehat{\Lambda^{\text{SF}}_D}(1) = 2.4119$, it has  a gradient of -8.9378, showing a strong negative incline of $\Lambda_D^\text{SF}$. 

\paragraph{$\text{RV}_D$ vs $\text{IVM}^\text{SF}_D$}
The regression (Fig. \ref{fig:rv-ivm-sf}) shows us a strong positive linear relationship between $\text{RV}_D$ and $\text{IVM}^\text{SF}_D$, suggesting $\sim10\%$ of runs survive indefinitely upon full systemic cooperation. However, the actual outcomes suggest it is possible to dramatically improve performance with less visibility. The highest point is $(0.75, 0.1333)$, where $\max \text{IVM}^\text{SF}_D(\text{RV}_D) = 0.1333$ and $\text{argmax}_{\text{RV}_D}\text{IVM}^\text{SF}_D = 0.75$, but $\widehat{\text{IVM}^\text{SF}_D}(0.75) = 0.0748$, overestimating by 0.0585. Either side of this are a pair of local minima with coordinates $(0.6667, 0.0167)$ and $(0.8333, 0.0125)$, respectively overestimating by 0.0495 and 0.0709. Before these, there exists a local maximum at $(0.6000, 0.0958)$, which is underestimated by 0.0392, and there exists a collection of adjacent values before it that are similarly underestimated. 

We see that $\text{IVM}^\text{SF}_D$ can be optimised in $M$, MH, and $H$, where
$M$ minimises visibility, saving expenses while allowing for inaccurate tuning to network diameter (as can be expected) through visibility reduction, which is a cost saving measure and thus appealing. MH gives the true maximum, but with a thin bidirectional margin of survivability, and so is risky. $H$ gives high returns with strong confidence and close survivability to the estimated range, but demands complete cooperation, which may be infeasible. 

Thus, according to IVM, it may be ideal to cap visibility to 0.6 to compromise between visibility expense, survivability, and reliability thereof.

\paragraph{$\text{RV}_D$ vs $\text{LTVM}^\text{SF}_D$} 
In this regression (Fig. \ref{fig:rv-ltvm-sf}), $\widehat{\text{LTVM}^\text{SF}_D}(1) = 0.1764$. Stretching a line from $(0, 0.0051)$ to $(1, 0.1764)$, we have an incline of 0.1713, showing a strong relationship between visibility and survival past the changepoint. However, the curve is not monotone. We see local minima at $(0.0733,-0.0147)$ and $(0.8519, 0.1144)$, and a local maximum at $(0.5620, 0.1569)$. Using the $M$, MH, $H$ separation, we also see a relative reduction of the MH region, with higher values in $M$ and $H$, including a global maximum in the data at $(0.4286, 0.5208)$ in $M$. With this we can explain the curvature of this graph versus the $\text{ITVM}^\text{SF}$ graph, since $M$ has been increased, MH has been reduced, and $H$ has increased.

The maximum of $(0.4286, 0.5208)$, while being a datapoint collected from a variety of networks and simulations, is abnormally prominent. We study now the regression of the data without it.

\paragraph{$(\text{RV}_D; \text{RV}_D \neq 0.4286)$ vs $\text{LTVM}^\text{SF}_D$}
This regression (Fig. \ref{fig:rv/rv0.4286-ltvm-sf}) is linear, and has the lowest $\sigma_{\hat{D}}$ of all measures for the scale free graphs, and $\widehat{\text{LTVM}^\text{SF}_D}(1; \text{RV}_D \neq 0.4286) = 0.1691$. The removal of $(0.4286, 0.5208)$ converts this into a first degree regression, but still has a reduced advantage of MH versus $M$ and $H$, as compared to $\text{LTVM}^\text{SF}_D$. In this case, $\langle M \rangle = 0.0854 < \langle \text{MH} \rangle = 0.1177 < \langle H \rangle = 0.1880$, suggesting one should maximise visibility. However, each set is bounded by minima and $M$ covers the widest range, so is the most predictable. These minima however are not so small as those in the $\text{LTVM}^\text{SF}$ graph, and also increase with $\text{RV}$. Thus, we see that, excluding $(0.4286, 0.5208)$, increasing visibility likely increases viability, inclusive of long-term and indefinitely surviving supply networks, but risks and expenses of achieving total cooperation may better justify visibility at the upper limits of $M$ and MH, which contain their respective maxima. 

Since $\sigma_{\hat{D}}$ is minimised for $(\text{RV}_D; \text{RV}_D \neq 0.4286)$ vs $\text{LTVM}^\text{SF}_D$, this may suggest that nonlinearities within our system are due to particular configurations of the supply network. To investigate this further, we repeat this analysis over Erd\H{o}s-R\'{e}nyi random graph generated supply networks.

\subsubsection{Random Networks}
For random graphs, short-lived supply networks almost totally dominate, with a mean of 0.9941 simulations of a given RV terminating before $B_D$. After this, a mean of 0.0035 simulations of a given RV are long-term but finite, occupying the long, flat, near zero region after each $B_D$. 

We also observe the prominent peak distribution is much more tightly clusters between 150 and 200 than that of the scale free diagrams.

\paragraph{$\text{RV}_D$ vs $\Lambda^\text{R}_D$}
In this case, short-lived simulations are tightly clustered around a single region. There is also minimal change in noisiness. We  examine the regression (Fig. \ref{fig:rv-lambda-sf}) over each identified RV$_D$ region.

There is a peak of $\widehat{\Lambda^\text{R}_D}$ in $L$ at  $(0.04766, 37.0053)$, which matches the peaks in $\widehat{IVM^\text{R}_D}$ and $\widehat{LTVM^\text{R}_D}$. 
Then there is a trough in ML at $(0.2136,  7.7134)$, matching those in $\widehat{IVM^\text{R}_D}$ and $\widehat{LTVM^\text{R}_D}$. 
$M$ contains three inflection points, with a peak at $(0.3351,11.103)$, a trough at $(0.5156, 6.0183)$, and another peak at $(0.6403, 7.7858)$. This motif corresponds to the singular peak in $M$ for $\widehat{IVM^\text{R}_D}$, inverting versus the high $(0.5, 0.0125)$ point in IVM$^\text{R}_D$.
There is then a trough in MH at $(0.7826,2.8318)$, corresponding to the trough in $\widehat{IVM^\text{R}_D}$ and peak in $\widehat{LTVM^\text{R}_D}$. 
Finally, there is a peak in $H$ at $(0.9518, 34.9296)$, corresponding to the peak in $\widehat{IVM^\text{R}_D}$ and trough in $\widehat{LTVM^\text{R}_D}$. 

\paragraph{$\text{RV}_D$ vs $\text{IVM}^\text{R}_D$}
At the detail of the quantities it shows, there are key differences.
The curve (Fig. \ref{fig:RVD_D1000}) shows no general inclines however there are three local maxima, at  $(0.0567, 0.0028)$, $(0.5254, 0.0053)$, and $(0.9822, 0.0040)$, and two local minima at $(0.2249, 0.0005)$ and $(0.8036, 0.0014)$. In this case, $M$ maximises $\widehat{\text{IVM}^\text{R}_D}$, MH minimises it, and $H$ receives a local maximum. If we also consider $L$ and ML,
then we see that ML contains a local minimum, and $L$ contains a local maximum, both of approximately similar height to their counters parts in MH and $H$, with only nominal skewness. This shows the distribution is reflected around $\text{RV}_D = 0.5$. 

Thus, there are 3 optimal strategies: constraining visibility to nonzero $L$, to the centre of $M$, or max of $H$, all while avoiding LM or MH. The non-zero performance of zero network coverage may be because of the fully random structure of this network. The $M$ range coverage performing the best is because agents in the system are supplied maximally relevant information for cooperation at the 0.5 relative visibility boundary. The full network coverage working well is due to capturing full systemic coverage. Results show that the outcomes for the random graph are not random themselves. We propose this is because of the systematic trimming action in the bankruptcy operation of our simulation. This may warrant further investigation. 

The maximum at $\text{RV}_D = 0.5$ also suggests that, according to $\text{IVM}^\text{R}_D$, agents in a full random system should have half network visibility, if they are to have visibility at all, or they can operate to a locally optimal level with only minimal visibility. It may be that because of the unstructured nature of the random graphs, information relevance does not linearly expand with visibility. Local regions are relevant according to the first peak, as is the whole of the network's dynamics according to the third peak, but the centre peak suggests a balancing of local relevance and systemic relevance leads to maximum viability. Again, however, the values encountered for the random graphs are very low, with nearly no incidence of indefinitely viable supply networks.

Reading Fig. \ref{fig:RVD_D1000}, the first peak in $L$ is when the neighbourhoods of $m$ and $r$ have each respectively visible. For the second peak in $M$, as half the diameter for a random network — while average path length grows logarithmically in Erd\H{o}s R\'enyi random graphs with size, traversal over the adjusted random graph is not yet clear. 

Adjacent to each peak is a trough, where node visibility has not harmonised with $m$ and $r$, instead putting more weighting on nodes in the network, which due to randomness, functions as noise in their internal predictive functions. The peak in $H$ is also when all nodes have both $m$ and $r$ visible.

\paragraph{$\text{RV}_D$ vs $\text{LTVM}^\text{R}_D$}
The curve (Fig. \ref{fig:RV_LTVM}) shows no clear directionality, however can be described according to inflection points and their relation with $L$, ML, $M$, MH, and $H$.

The first peak of the curve occurs in $L$ at $(0.0589, 0.0174)$, likely skewed by $(0.1667, 0.0167)$,  closely matching the $\text{IVM}^\text{R}$ regression but amplified to a global maximum. 
The first trough is at  $(0.2554, 0.0002)$, contained in ML, matching the zeros that span it. Another peak exists at $(0.7488, 0.0108)$, in MH, preceded by steady growth in $M$. Finally, there is a trough with negative $\widehat{\text{LTVM}^\text{R}_D}(\text{RV}_D)$ in $H$ at $(0.9492, -0.0035)$, likely due to overshoot while still having $\widehat{\text{LTVM}^\text{R}_D}(1)  = 0.0052 = \text{LTVM}^\text{R}_D(1)$.

Discussing these values, we first note that $\max_{\text{RV}_D} \text{LTVM}^\text{R}_D = 0.0167$, and the rate of $\text{LTVM}^\text{R}_D$ viability is to the same order of magnitude as $\text{IVM}^\text{R}_D$, and restricted to 2\% of possible networks as viable per RV. In the structured, scale free case, $\text{LTVM}^\text{SF}_D$ is an order of magnitude larger than $\text{IVM}^\text{SF}_D$. Thus, random graph simulations with RV in $M$, if viable, are indefinite and rare, while those with RV in $L$ or MH, if viable, are long-term but finite, and to the same frequency. 

This can be seen in how Fig. \ref{fig:RV_LTVM} shows a distinct topology from Fig. \ref{fig:RVD_D1000}, where the vertical proportion is approximately the same, however, there exists some incidence of long-term viability for extremal RV.

\bibliographystyle{abbrv} 
\bibliography{references}

\end{document}